\documentstyle[pre,aps,multicol]{revtex}

\def\lsim{\buildrel {\textstyle <}\over {_\sim}}

\tighten
\input epsf

\begin{document}
\draft

\title{\Large \bf  
Successive transitions and intermediate chiral phase \\
in a superfuilid $^3$He film}

\author{\bf Hikaru Kawamura}

\address{Faculty of Engineering and Design, 
Kyoto Institute of Technology, Kyoto 606-8585, Japan}

\maketitle

\begin{abstract}
Superfluidity ordering of thin $^3$He films is studied by
Monte Carlo simulations based on a two-dimensional
lattice spin model with 
$Z_2\times U(1)\times SO(3)$ symmetry. Successive phase transitions
with an intermediate `chiral' phase, in which
the {\bf $l$}-vector aligns 
keeping the phase of the condensate disordered,
is found.  
Possible experiments to detect the successive transitions
are discussed.
\end{abstract}


\begin{multicols}{2} \narrowtext

It is now well established that the superfluidity transition 
of $^4$He films belongs to the Kosterlitz-Thouless (KT)
universality class 
governing a variety of 
two-dimensional (2D) systems with $U (1)$ symmetry [1].
By contrast, the nature of the 2D phase transition 
of its isotope, {\it i.e.\/},  the superfluidity transition of
$^3$He films of thickness of order the zero-temperature coherence
length,  $0.01\mu m\lsim \xi _0\lsim 0.05\mu m$, 
has not been well understood.
Although the complex structure in the order parameter 
of $^3$He [2]
makes its transition behavior  an intriguing problem, 
the experimental observation of  
the possible nontrivial critical behavior of  {\it bulk\/} $^3$He
is difficult, 
since, due to its long 
coherence length, 
the critical regime is limited
to a narrow temperature range close to 
$T_c$, {\it i.e.\/}, 
$t\equiv
\mid (T-T_c)/T_c\mid \lsim t^*\approx 10^{-6}$ [3].
By contrast, in the case of thin films, 
enhanced effects of fluctuations due to the low
dimensionality give rise to much wider critical regime, as wide as
$t^*\approx 10^{-1}\sim 10^{-2}$ [4].
Hence, there appears to be a good possibility
to experimentally observe the novel transition behavior originated
from the complex order parameter.

There have been many  experimental works
on the superfluidity ordering of $^3$He in restricted
geometries including  cylindrical pores [5,6] and  thin flat films 
[7,8].
Indeed, several experiments reported the observation of 
novel ordering behaviors,
even  a signature of new phase [6,7].
Experimentally, one challenging
problem is to make measurements on
a sample which can be regarded as a good 2D system, 
{\it i.e.\/}, a thin flat film of uniform
thickness of order $0.1\mu m$ or less 
on sufficiently specular
boundaries. 
For such an ideal 2D system, theoretical calculation suggests
that the stable  superfluidity
state is that of the $A$ phase  down to zero pressure [9],
which has also been confirmed experimentally [8].
The order parameter can be described as a triad of
three orthogonal unit vectors ${\bf m}$, ${\bf n}$ and ${\bf l}=
{\bf m}\times {\bf n}$, where the ${\bf l}$-vector is kept
perpendicular to the surface. 

Some time ago, Stein and Cross argued that
such superfluidity state with the A-phase structure 
possessed the novel twofold Ising-like degeneracy, 
according as the ${\bf l}$-vector
was either {\it parallel\/}
or {\it antiparallel\/} with the surface normal [10].
This $Z_2$ symmetry coexists with
the standard $U(1)$ gauge symmetry associated with the 
phase of the condensate. The latter symmetry corresponds to
a continuous
rotation of the ${\bf m}$- and ${\bf n}$-vectors  around the 
${\bf l}$-vector.
Stein and Cross then argued that this novel $Z_2$ symmetry might 
lead to an Ising-like transition which might ``compete'' 
with the standard
KT transition associated with the $U(1)$ gauge symmetry, although
the detailed nature of the transition was not specified.

Since then, 2D phase transitions associated 
with  the $Z_2\times U(1)$ 
symmetry have been studied   in a different area,
{\it i.e.}, in the context of
frustrated 2D {\it XY\/} systems such as
the triangular-lattice {\it XY\/} antiferromagnet [11]
or the Josephson-junction
array in a magnetic field [12]. In these problems, the 
$Z_2$ Ising-like degree of
freedom has been called `chirality' [13].
While these studies have established the 
occurrence of a phase transition with a sharp specific-heat
anomaly, 
there still remains some controversy 
as regards whether the
$Z(2)$ and $U(1)$ degrees of freedom order simultaneously, or at
two close but distinct temperatures.
Meanwhile, some differences  exist between these 
frustrated {\it XY\/}systems and
$^3$He films.
Particularly important point neglected in the previous analysis [10]
may be  the fact that
$^3$He has an additional internal symmetry, $SO(3)$, 
associated with the nuclear spin degree of freedom of the condensate. 

In the present Letter, I  perform a numerical study of the
superfluidity ordering of $^3$He films in order to clarify
possible thermodynamic phases and
the transition behavior between them. For that purpose, 
I introduce a 2D lattice spin (pseudospin) model which possesses
the expected full symmetry of the order parameter,
$Z_2\times U(1)\times SO(3)$, and perform 
Monte Carlo simulations.
It is found that the model exhibits two successive transitions
with a `chiral' intermediate phase.

The order parameter describing the 2D superfluidity of 
$^3$He may be described
by a $3\times 2$ tensor variable $A_{\mu j}$,
\begin{eqnarray}
A_{\mu j}=A d_\mu (m_j+in_j), \nonumber
\\
\mid {\bf d}\mid = \mid {\bf m}\mid =\mid {\bf n}\mid =1,\ \ 
{\bf m}\cdot {\bf n}=0,
\end{eqnarray}
where $\mu =x,y,z$ refers to the spin component,  $j=x,y$ refers 
to the
real-space coordinate (the film surface is taken to be the 
$xy$-plane),
${\bf d}$ is a three-component unit vector in spin space
representing the spin state of the condensate,  while ${\bf m}$ and
${\bf n}$ are mutually orthogonal two-component
unit vectors in  real space
representing the orbital state of the condensate. The Ising-like
variable, or the chirality,
is defined by 
$\tau ={\rm sign} (l_z)=m_xn_y-m_yn_x$, where
$\tau =\pm 1$ represents  the ${\bf l}$-vector pointing either up
or down along the surface normal. 

In deriving an appropriate model Hamiltonian, I start with 
the standard Ginzburg-Landau free energy in  the London limit,
\begin{eqnarray}
{\cal H}_{GL}\approx K(\mid \nabla {\bf A}_\mu \mid ^2 + 
\delta \mid {\rm div}{\bf A}_\mu \mid ^2),
\end{eqnarray}
where ({\bf A}$_\mu $)$_i$=$A_{\mu i}$.  
Although there generally exists a spatially-anisotropic 
gradient term (the second term of Eq.(2)), 
renormalization-group $\epsilon =4-d$ expansion analysis showed
that such spatial 
anisotropy was irrelevant at the superfluidity
transition, {\it i.e.\/}, $\delta \rightarrow 0$ upon 
renormalization [3]. Hence, for simplicity, 
I drop the second term here.
Neglecting the weak dipolar interaction of order $10^{-1}\mu $K, 
restricting the space to 2D, parametrizing the ${\bf m}$-vector
as ${\bf m}=(\cos \theta ,-\tau \sin \theta)$ where $0\leq \theta 
<2\pi$ is the phase angle of the condensate,  
and discretizing the continuum into the lattice, one  obtains
\begin{equation}
{\cal H}=-J\sum _{<ij>}(1+\tau_i\tau_j)\cos (\theta_i-\theta_j)
{\bf d}_i\cdot {\bf d}_j,
\end{equation}
where $J>0$ is a coupling constant and
the summation is taken over all nearest-neighbor pairs on
the square lattice. The Hamiltonian (3) can be viewed as a 
coupled Ising-{\it XY\/}-Heisenberg
model with  $Z_2\times U(1)\times SO(3)$ symmetry, in
which $\tau _i=\pm 1$ is an Ising variable [$Z_2$], 
$\theta _i$ being an angle
variable [$U(1)$] of the {\it XY\/}-pseudospin 
${\bf p}_i=(\cos \theta _i,\sin \theta _i)$,
and {\bf d}$_i$ is a Heisenberg variable [$SO(3)$].
Note that the Hamiltonian (3) also possesses the 
{\it local\/} symmetries under the transformations;
(i) ${\bf p}_i\rightarrow \tau_i{\bf p}_i$;  (ii) 
${\bf d}_i\rightarrow \tau_i{\bf d}_i$;  (iii) 
${\bf p}_i\rightarrow \pm{\bf p}_i$ and
${\bf d}_i\rightarrow \pm{\bf d}_i$.

The ordering of the ${\bf l}$-vector, or of the chirality, 
can be probed via the 
Ising-magnetization,  $\tau ={1\over N}\sum _i\tau _i$
($N$ is the total number of lattice sites), by
calculating the average chirality
and the associated Binder ratio,
\begin{equation}
\bar \tau =<\tau^2>^{1/2},\ \ 
g_\tau ={1\over 2}(3-{<\tau ^4>\over <\tau ^2>^2}).
\end{equation}
Since the correlation functions associated with the
phase variable $\theta $, or the {\it XY\/}-pseudospin 
${\bf p}$, are not invariant under the above local transformations 
and vanish trivially,  
one needs to calculate the
``local-gauge-invariant'' quantity to study the phase ordering.
An appropriate quantity is a second-rank symmetric traceless tensor 
$P_{i\mu \nu}={1\over \sqrt 2}(p_{i\mu}p_{i\nu}-{1\over 2}
\delta _{\mu \nu})$, or equivalently,  a new {\it XY\/}-pseudospin 
variable
with the doubled phase angle, ${\bf p}'_i=
(\cos 2\theta_i, \sin 2\theta _i)$. 
Via the {\it XY\/}-magnetization, ${\bf p}'={1\over N}
\sum _i{\bf p}'_i$, the associated Binder ratio is defined by
\begin{equation}
g_\theta =2-{<{\bf p}'^4>\over  <{\bf p}'^2>^2}.
\end{equation}
Likewise, the ordering of the
${\bf d}$-vector is probed via
$D_{\mu \nu}={1\over N}\sum _iD_{i\mu \nu}$, where a
second-rank symmetric traceless tensor is defined by 
$D_{i\mu \nu } 
={1\over \sqrt 3}(d_{i\mu }d_{i\nu }-{1\over 3}\delta _{\mu \nu})$.
The corresponding Binder ratio is given by
%
%
\begin{equation}
g_{{\bf d}}=\frac {1}{2}(7-5\frac{<D^4>}{<D^2>^2}),
\ \ \ \ D^2\equiv \sum _{\mu \nu}D_{\mu \nu}^2.
\end{equation}

Monte Carlo 
simulations are performed based on the standard 
single-spin-flip Metropolis method.
The lattice contains
$N=L^2$ sites with $L=20,30,40$ and 60 with periodic boundary conditions.
Typically, I generate total of $10^5$ Monte Carlo steps per spin (MCS)  
at each temperature, and $(2.5\sim 10)\times 10^5$
MCS in the transition region.

The calculated specific heat is shown in Fig.1. A sharp peak
is observed at $T/J=0.754\pm 0.002$.
As is shown in the inset, 
the peak grows slowly with increasing $L$, indicative of a continuous 
transition characterized by the exponent
$\alpha \sim 0$.

\bigskip\centerline{
\epsfysize=6.5cm
\epsfxsize=9cm
\epsfbox{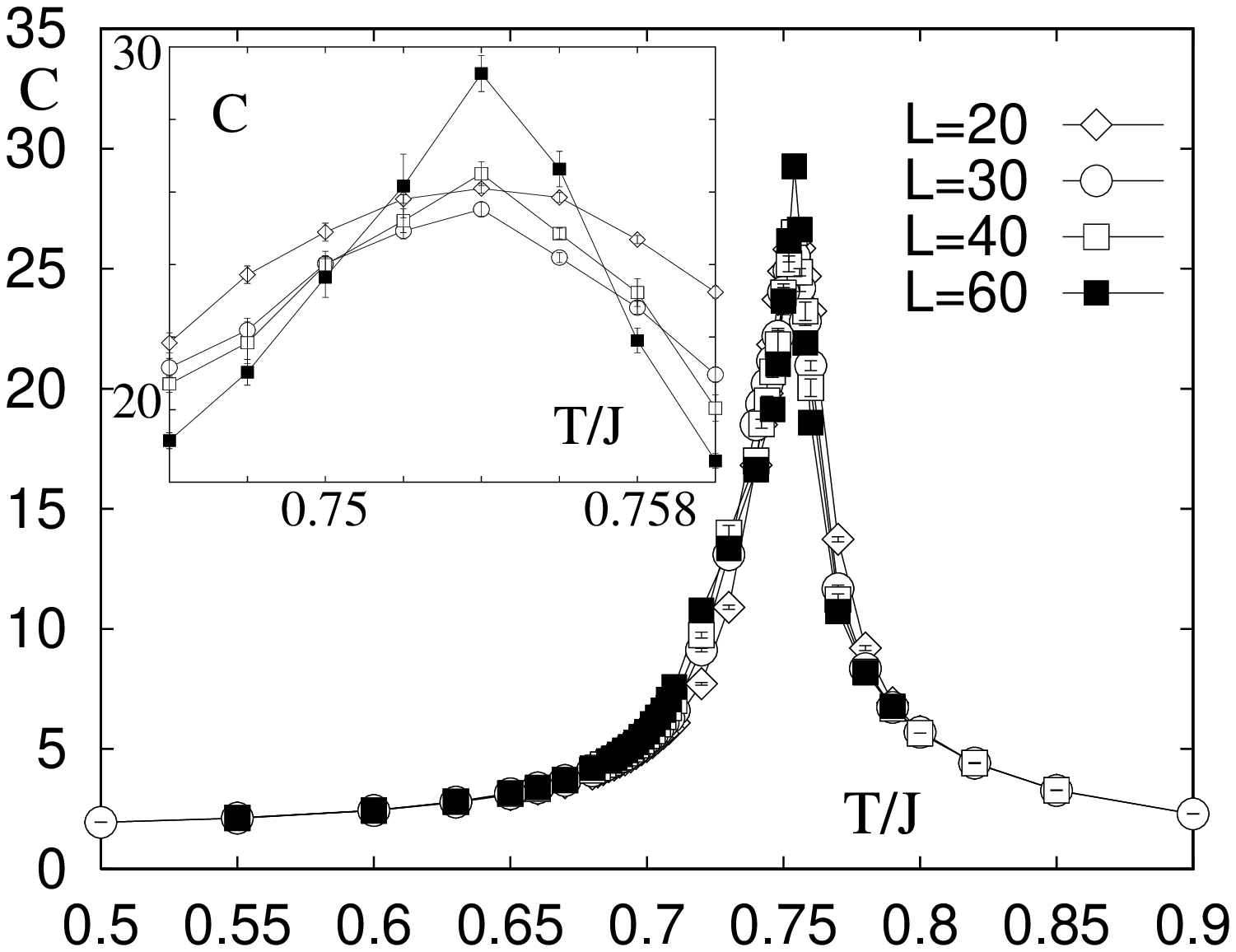}}
\noindent
FIG.1~ The temperature and size dependence of the 
specific heat per site. The inset represents a magnified view of the
transition region.

\bigskip

The calculated Binder ratios (4)-(6)
are shown in Figs.2a-c, respectively.
As can be seen from Fig.2a, $g_\tau$ for different $L$ cross 
at a temperature 
which coincides with the specific-heat-peak temperature.
This indicates
the occurrence of
a continuous transition 
at $T_{c1}/J=0.753\pm 0.001$ into the phase with a finite 
chiral long-range order, $\bar \tau >0$, 
accompanied by a sharp specific-heat 
anomaly.  
The behavior of
$\bar \tau$ (not shown here) 
turns out to be fully consistent with this.
By contrast, as can be seen from Figs.2b and c,
$g_\theta $ and $g_{{\bf d}}$ for various $L$ 
donot cross or merge at $T_{c1}$, constantly decreasing with 
increasing $L$ around $T_{c1}$, 
a behavior characteristic of a disorder phase.
Hence, the state just below $T_{c1}$ is a pure chiral 
state with only a 
chiral long-rage  order while the phase and the spin are
kept disordered.

At a lower temperature $T_{c2}/J=0.692\pm 
0.04$,   $g_\theta $ for various $L$
merge, and below $T<T_{c2}$, continue to stay on a common curve: 
See Fig.2b.
Such a behavior is

\bigskip\centerline{
\epsfysize=6.5cm
\epsfxsize=9cm
\epsfbox{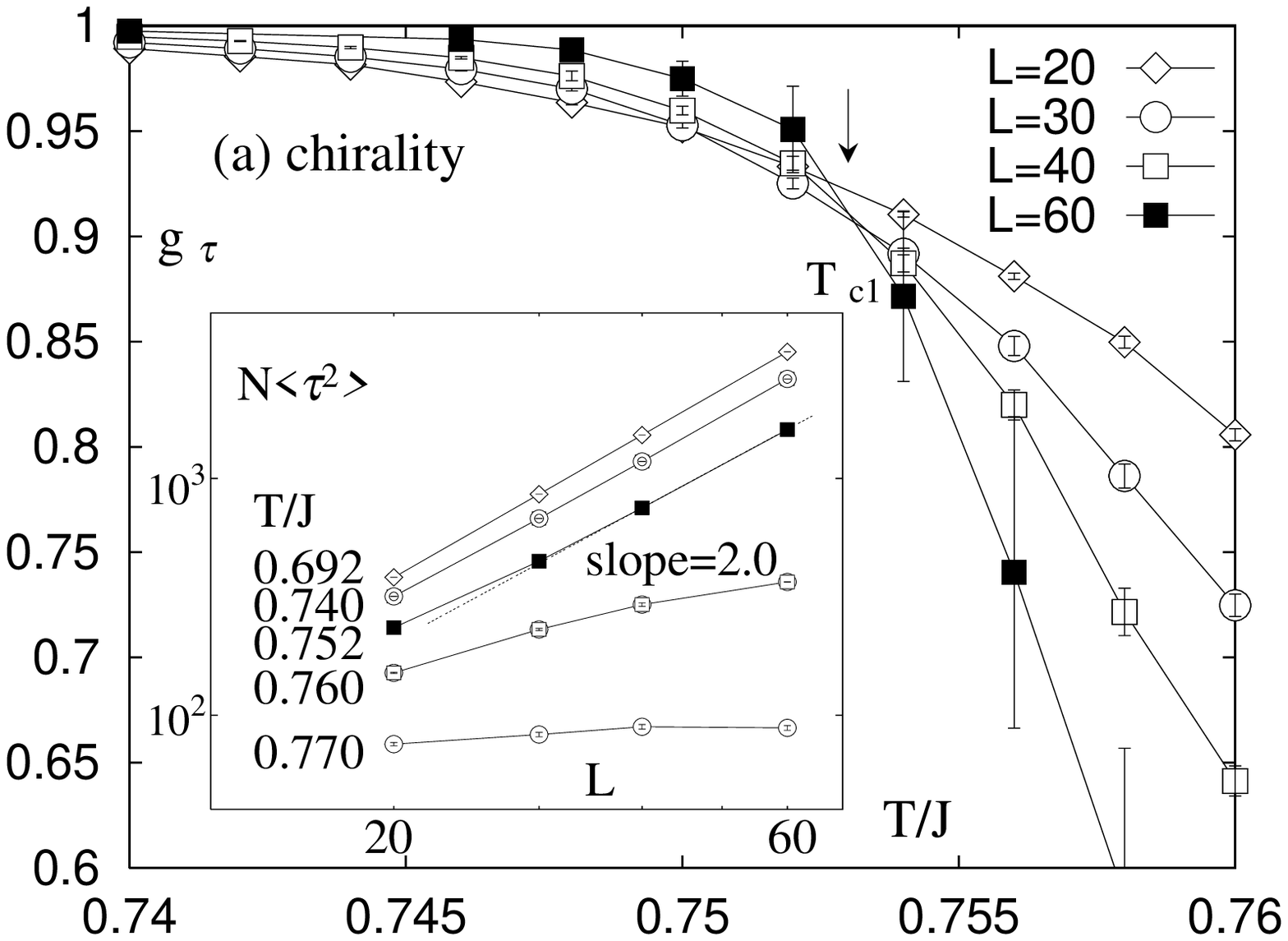}}

\smallskip\centerline{
\epsfysize=6.5cm
\epsfxsize=9cm
\epsfbox{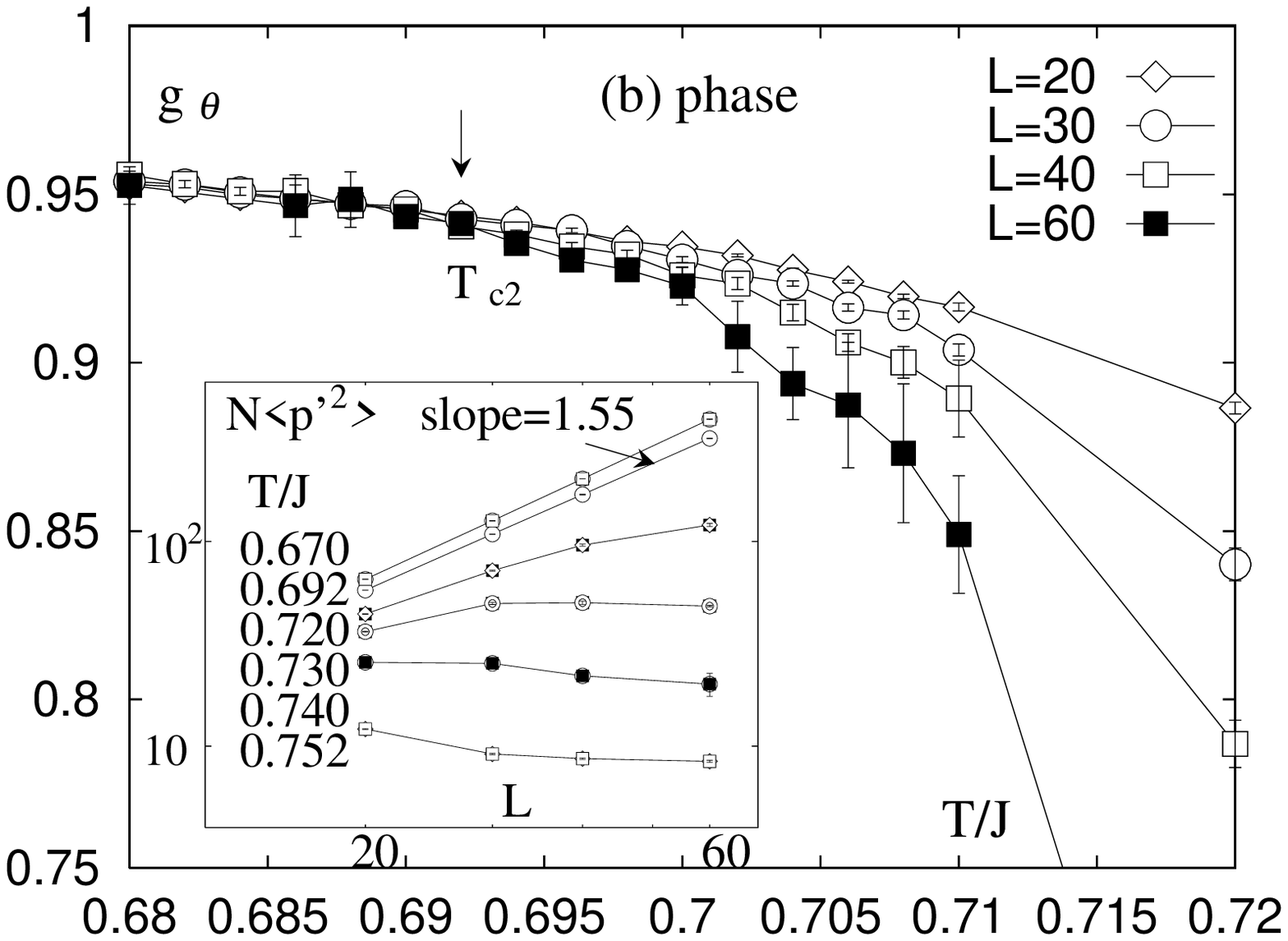}}

\smallskip\centerline{
\epsfysize=6.5cm
\epsfxsize=9cm
\epsfbox{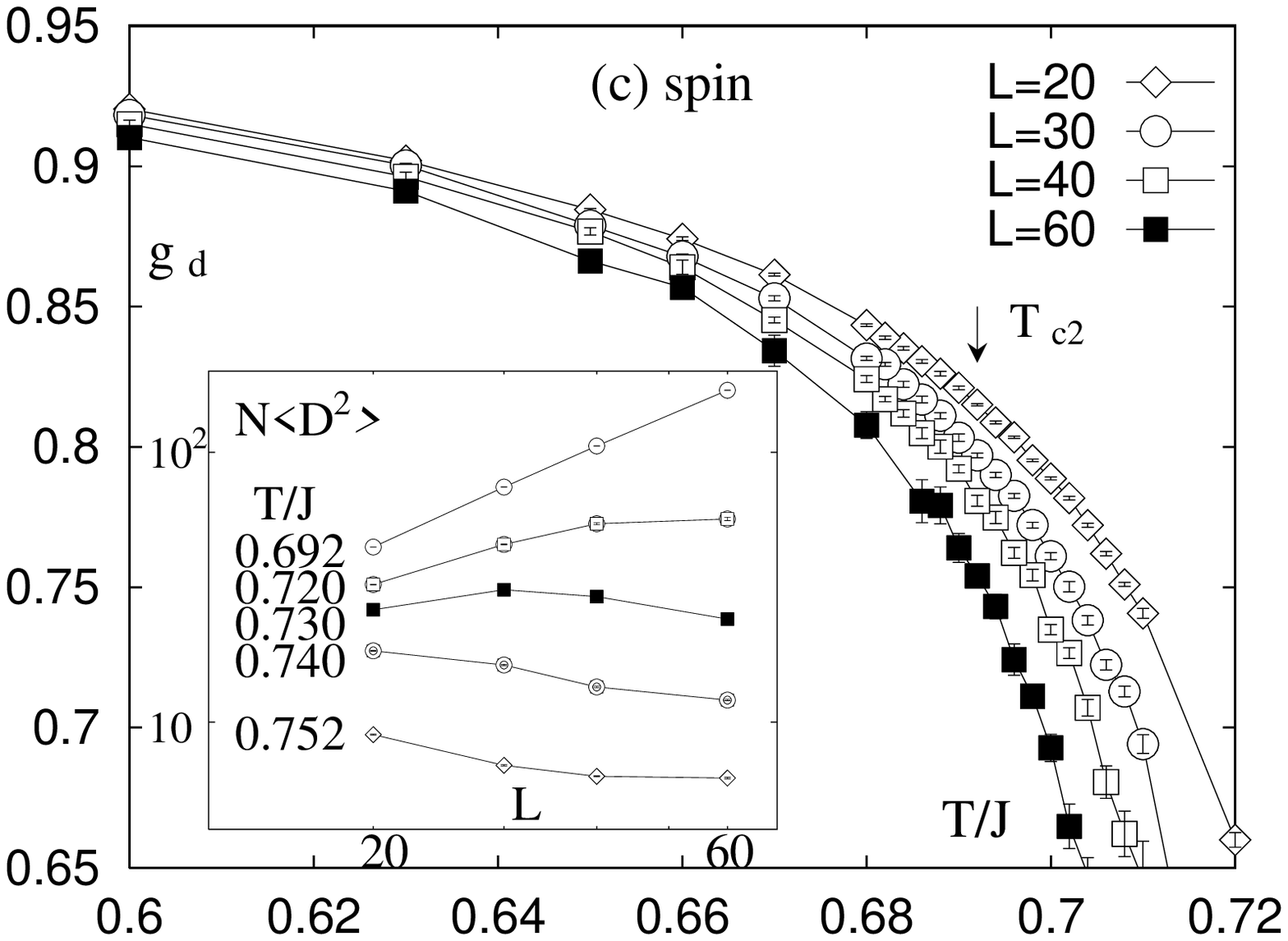}}

\noindent
FIG.2~ The temperature and size dependence of the Binder 
ratios in the transition region; 
(a) of the $\tau $-variable (chirality); (b) of the 
${\bf p'}$-variable (phase); (c) of the {\bf d}-variable (spin). 
Note the difference in the temperature scale in each figure. 
The insets show log-log plots of the $L$ dependence of the ordering
susceptibilities, $N<\tau ^2>$ (a), $N<{\bf p'} ^2>$ (b) and
$N<D^2>$ (c).  In the ordered phase with a nonzero order parameter,
the data for larger $L$ should lie on a straight line 
with a slope equal to two, 
while in the disordered phase the data should exhibit a downward
curvature staying finite even in the $L\rightarrow \infty $ limit.
At criticality and in the KT-like phase, the data should lie on a
straight line with a slope $2-\eta $, which is less than two.

\bigskip
\noindent
characteristic of the KT-like transition
in which the low-temperature phase is a critical phase with
algebraically-decaying correlations. Thus, the quasi-long-range order
of the $U(1)$ phase
is realized below $T_{c2}$. The absence of any appreciable anomaly in
the specific
heat around  $T_{c2}$ is also consistent with such 
KT-like transition.
By contrast, $g_{{\bf d}}$ monotonically 
decreases with increasing $L$
around $T_{c2}$, suggesting that the ${\bf d}$-vector 
remains 
disordered even at $T<T_{c2}$. This observation is consistent with the
fact that the
${\bf d}$-vector is a Heisenberg variable carrying 
the $SO(3)$ symmetry, and that the 2D Heisenberg model orders 
only at $T=0$. I have also checked that the behaviors 
of the  order parameters corroborate the above 
conclusion. As an example, the $L$ dependence of the  
ordering susceptibilities, $N<\tau ^2>$, $N<{\bf p}'^2>$ and
$N<D ^2>$, are shown on log-log plots in the insets of Fig.2.

Thus, the present model exhibits two successive transitions at
$T=T_{c1}$ and $T_{c2}$, where 
$T_{c1}\simeq 0.753J>T_{c2}\simeq 0.692J$.
The intermediate phase is a pure chiral state in which
only the $Z_2$ chirality, representing the direction of the
${\bf l}$-vector, aligns.
In the low-temperature phase, in addition to the chiral long-range 
order, the $U(1)$ phase of the condensate
exhibits the KT-type algebraic order leading to a true 
superfluidity.
Meanwhile, the ${\bf d}$-vector representing the spin state of the
condensate 
remains disordered at any finite temperature.
Although in real films the dipolar interaction neglected here 
gives rise to additional interaction, 
it is several orders of magnitude weaker than $J$ [2]  
and the ordering of
the ${\bf d}$-vector, if any, 
should occur at a temperature much lower than $T_{c1}$ and $T_{c2}$.

To identify the universality class of each transition,
standard finite-size scaling
analysis is made. 
The estimated chirality exponents at $T=T_{c1}$, $\nu =1.0\pm 0.1$ and 
$\beta/\nu =0.13\pm 0.01$, 
agree well 
with the values of the 2D Ising model, 
and I conclude that
the upper transition belongs to the standard 2D Ising universality 
class. 
As regards the lower transition, the 
decay-exponent $\eta _\theta $ associated with the correlation function of the
${\bf p}'$-vector at $T=T_{c2}$, 
$<{\bf p'}({\bf r})\cdot {\bf p'}({\bf 0})>
\approx r^{-\eta _\theta }$,  
is estimated from the log-log plot of 
$N<{\bf p}'^2>$ versus $L$, 
yielding $\eta _\theta =0.45\pm 0.03$: See the inset of Fig2b.
The estimated $\eta _\theta $
is considerably larger than the standard KT value, $\eta =0.25$,
suggesting that the lower transition is {\it not\/} of the
standard KT universality. It should be noticed, however, that
the spatial anisotropy neglected in deriving
our model Hamiltonian (3) might possibly 
affect the value of $\eta _\theta $.

In the present model,
the intermediate phase is realized over an
appreciable temperature range, 
in sharp contrast to the case of frustrated {\it XY\/} models where
$T_{c1}$ and  $T_{c2}$ are identical or very close [11,12].
This difference is
probably caused by the $SO(3)$ degree of freedom associated with the 
{\bf d}-vector. Namely, although the {\bf d}-vector itself does
not order at any $T>0$, its existence considerably 
affects the effective interaction between
the chiral and phase variables. 
This may be seen by deriving an effective Hamiltonian 
${\cal H}_{eff}$ written
in terms of the $\tau $- and $\theta $-variables, by integrating out
the ${\bf d}$-variable. 
At sufficiently high temperatures, one can explicitly carry out this 
procedure to get
\begin{eqnarray}
{\cal H}_{eff}/k_BT\approx -\sum _{<ij>}\{K_1(1+\tau_i\tau_j)
\cos (2\theta_i-2\theta_j) \nonumber \\
+K_2\tau _i\tau _j\}, 
\end{eqnarray}
where one has $K_1=K_2$. Indeed, a numerical study in Ref.[14]
suggested that in the case of
$K_1=K_2$ the model exhibited two successive transitions with an
intermediate chiral phase consistent with the present result, whereas
the frustrated {\it XY\/} models corresponded to $K_2\simeq 0$ where
the model (7) was estimated to exhibit a single transition only.

Finally, I discuss experimental implications of the 
obtained results.
Main conclusion of the present work is that successive transitions and an 
intermediate phase are likely to be realized in thin films of
$^3$He.
While the clearest sign of the upper transition is
a logarithmically divergent 
anomaly in the specific heat, it might practically be
difficult to get the necessary experimental sensitivity because 
specific-heat measurements usually require a considerable mass of
sample.
Some anomaly
might be detectable at $T=T_{c1}$, however,  in the quantities like 
thermal-expansivity, optical properties, and
sound-velocity or attenuation (particularly, the third sound [15]). 
By contrast, 
the NMR shift, which has widely been used in identifying the 
bulk superfluidity
transition,
is blind to the
distinction between ${\bf l}$ and $-{\bf l}$, and may not be a very good
indicator of the transition. Rather, the shift should 
set in
somewhat {\it above\/} $T_{c1}$, by an amount of order $t^*$,
when the  short-range order is 
developed and the {\bf l}-vector stays perpendicular to the surface.

Although the intermediate phase is {\it not\/} a true
superfluid, it is expected to be a low-dissipation state
distinguishable from the high-dissipation normal state above $T_{c1}$.
Interestingly, the observation of a low-dissipation state was
reported in the literature [5,6].
By contrast, the lower transition is characterized by the onset 
of a true superfluidity, and is
detectable by the standard torsional-oscillator measurements.
While the areal superfluid density $\rho _s$ is
expected to show a non-KT jump at $T=T_{c2}$, 
$\rho _s/T_{c2}=8m^2k_B/(\pi \eta _\theta \hbar ^2)$ 
($m$ is the mass of $^3$He atom),
its absolute value $\rho _s/T_{c2}\sim 
17.6\times 10^{-12}$g/(cm$^2$mK) is small 
reflecting the low $T_c$ of $^3$He, and
experimental observation of the jump itself might be difficult.

Anyway, in the 2D superfluidity ordering of $^3$He, 
new interesting physics  different from that of helium four
is expected. Further experimental
studies are encouraged.

I thank K. Kono and K. Shirahama for useful discussions. 
The numerical calculation was performed on the FACOM
VPP500 at the supercomputer center, ISSP, University of Tokyo.

\end{multicols}\widetext

\end{document}